\documentclass[a4paper,11pt]{article}
\pdfoutput=1 
\usepackage[style=ieee,backend=biber,minnames=1,maxnames=1,maxcitenames=1]{biblatex}

\usepackage{jinstpub} 
\usepackage{siunitx}
\usepackage[latin1,utf8]{inputenc}
\usepackage{subfig}
\usepackage{wrapfig}

\newcommand{\pitch}[2]{
\SIrange[range-phrase=\times]{#1}{#2}{\micro\meter\squared}
}
\newcommand*\chem[1]{\ensuremath{\mathrm{#1}}}

\title{\boldmath Pixel detector R\&D for the Compact Linear Collider}

\author[a,1]{M.~Benoit\note{Corresponding author.}}
\affiliation[a]{D\'epartement de Physique Nucl\'eaire et Corpusculaire
  (DPNC), Universit\'e de Gen\`eve, 24 Quai Ernest Ansermet 1211 Gen\`eve 4, Switzerland}
\emailAdd{Mathieu.Benoit@cern.ch}

\collaboration[c]{on behalf of CLICdp Collaboration}

 \abstract{The physics aims at the proposed future CLIC high-energy linear $e^+ e^-$ collider pose challenging demands on the performance of the detector system. In particular the vertex and tracking detectors have to combine precision measurements with robustness against the expected high rates of beam-induced backgrounds. A spatial resolution of a few microns and a material budget down to 0.2\% of a radiation length per vertex-detector layer have to be achieved together with a few nanoseconds time stamping accuracy. These requirements are addressed with innovative technologies in an ambitious detector R\&D programme, comprising hardware developments as well as detailed device and Monte Carlo simulations based on TCAD, Geant4 and Allpix$^2$. Various fine pitch hybrid silicon pixel detector technologies are under investigation for the CLIC vertex detector. The CLICpix and CLICpix2 readout ASICs with \SI{25}{\micro\meter} pixel pitch have been produced in a \SI{65}{\nano\meter} commercial CMOS process and bump-bonded to planar active edge sensors as well as capacitively coupled to High-Voltage (HV) CMOS sensors. Monolithic silicon tracking detectors are foreseen for the large surface ($\approx$ \SI{140}{\meter\squared}) CLIC tracker. Fully monolithic prototypes are currently under development in High-Resistivity (HR) CMOS, HV-CMOS and Silicon on Insulator (SOI) technologies. The laboratory and beam tests of all recent prototypes profit from the development of the CaRIBou universal readout system. This paper presents an overview of the CLIC pixel-detector R\&D programme, focusing on recent test-beam and simulation results.}


\begin{document}
\titlepage
\maketitle
\flushbottom

\section{Introduction}

The CLIC high-energy linear $e^+ e^-$ collider, under development by international collaborations hosted by CERN, is based on the novel two-beam acceleration method, operating with large gradients of \SIrange{70}{100}{\mega\V\per\m} in a normal conducting accelerating structure. It is proposed to be operated in a staged scenario with respectively \SI{380}{\GeV}, \SIlist{1.5;3}{\TeV} centre-of-mass energy \cite{robson_compact_2018}. The CLIC beam is composed of trains that consist of 312 bunches separated by \SI{500}{\ps} with a \SI{50}{Hz} train repetition rate. In the CLIC vertex and tracking detector, a high rate of incoherent pairs and $\gamma\gamma \xrightarrow~ $ hadrons generated from beam-beam interactions. 

\begin{figure}[h]
\begin{minipage}[b]{0.5\textwidth}
\begin{center}
  \includegraphics[width=0.5\textwidth]{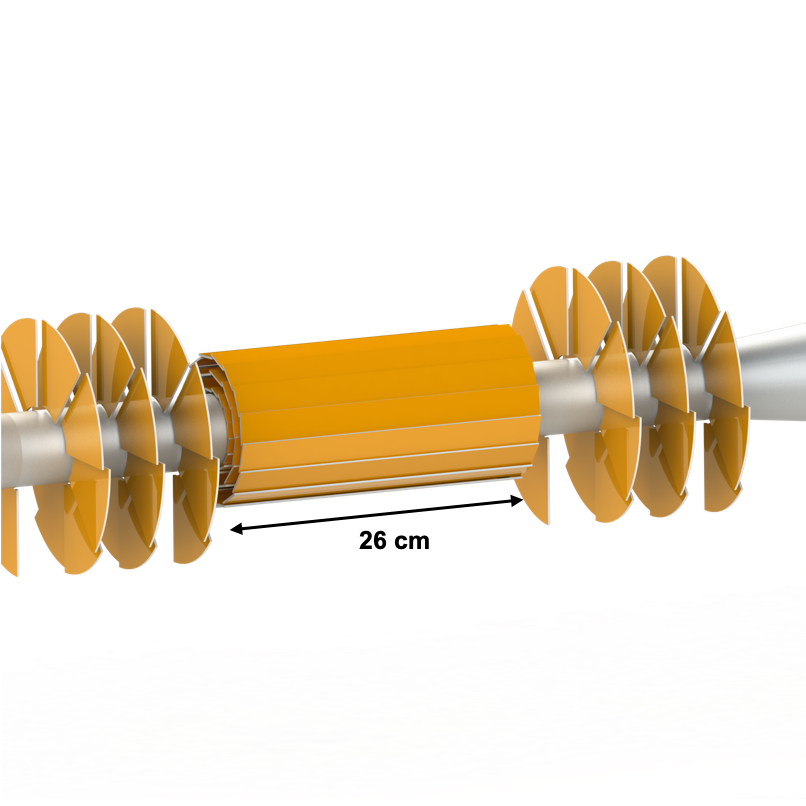}
  \caption{Layout of the CLIC vertex detector.}
  \label{vertex}
  \end{center}
 \end{minipage}
  \begin{minipage}[b]{0.5\textwidth}
\begin{center}
    \includegraphics[width=0.6\textwidth]{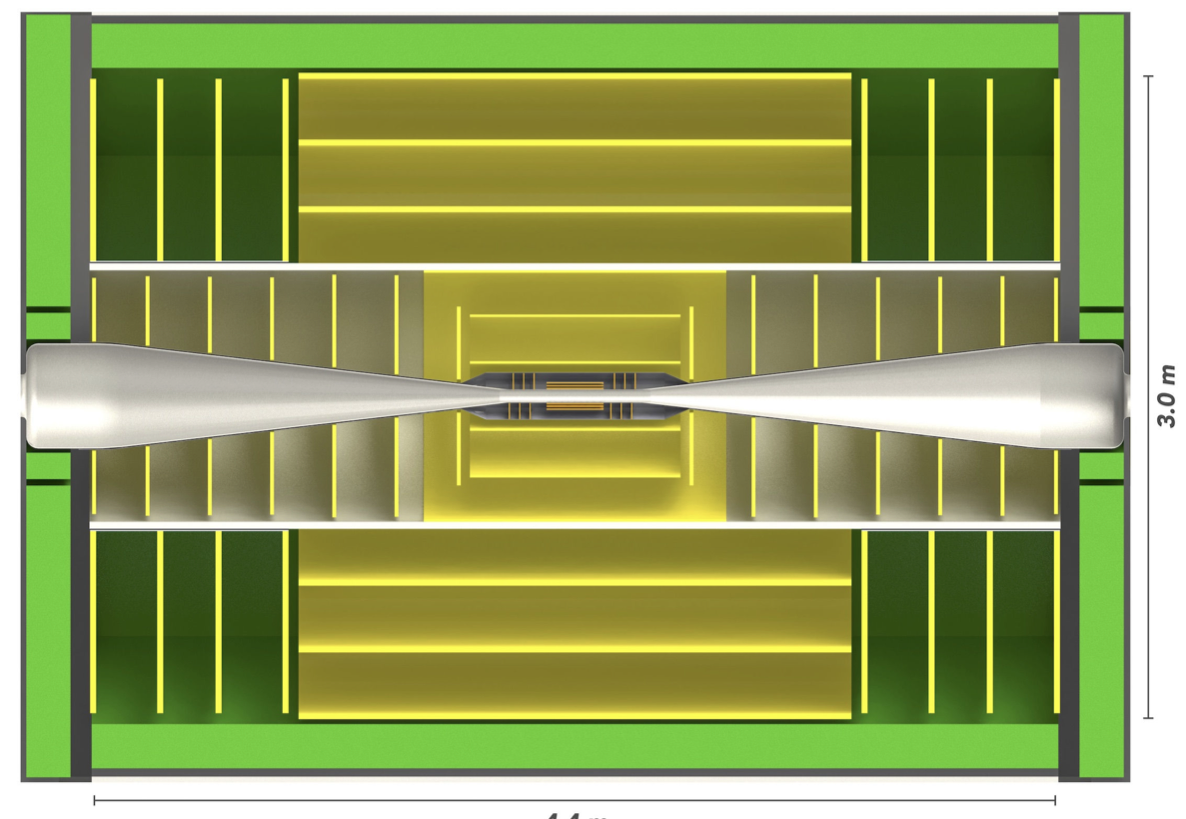}
  \caption{Layout of the CLIC tracking detector.}
  \label{tracker}
  \end{center}
  \end{minipage}
\end{figure}

The CLIC detector aims at performing high precision measurements of standard and beyond standard model physics. This imposes challenging performance requirements on the  detectors.  The vertex detector, illustrated in Figure \ref{vertex}, consists of barrel and disks directly surrounding the beam pipe, covers a surface of \SI{0.84}{\meter\squared} and is composed of pixel detectors with a pitch of $\pitch{25}{25}$ for an expected single point resolution of \SI{3}{\micro\meter}. The Tracking detector, illustrated in Figure \ref{tracker}, is composed of pixelated silicon detectors with a pixel pitch of $\pitch{25}{1000}$ and covering a large area of \SI{140}{\meter\squared}. In both detectors, in order to cope with the occupancy and help with pattern recognition, a timing resolution of \SI{5}{\nano\second} is required. To meet the performance requirements of CLIC, the material budget allowed for each of the six layers and seven disks of the pixel detector is \SI{0.2}{\percent~X_0 \per layer} and \SIrange{1}{2}{\percent~X_0 \per layer} for the five layers of the Tracker, including local support and services.

Taking advantage of the low duty cycle of the accelerator, the vertex and tracking detector are operated using power pulsing. The readout and front-end electronics are turned off when not in use, in order to reduce the power consumption to \SI{50}{\milli\watt\per\cm\squared} for the vertex detector and below \SI{150}{\milli\watt\per\cm\squared} for the Tracking detector. This reduction in power consumption allows for the use of air cooling which reduces the need for material for active cooling in the vertexing volume. The radiation damage estimated in the vertex detector is of less than \SI{1e11}{n_{eq}\per \cm\squared \per year} and \SI{1}{\kilo Gy \per year} and is considered negligible in the Tracking detector.

\section{CLICdp tools for pixel detector prototype characterisation and simulation }

The CLICdp  collaboration studies the available pixel detector technologies, evaluates their tracking performance and identifies promising technologies for the construction of the CLIC tracking and vertex detectors. In order to carry out the characterization and modeling of the prototypes realized in these technologies, a set of tools have been developed. 

To extract the single-point resolution, particle detection efficiency and timing resolution of the prototypes, the Timepix3 telescope \cite{alipour_tehrani_test-beam_2017,dallocco_timepix3_2018} has been constructed. To provide a versatile platform for the development of a readout system for each of the prototypes, the CaRIBOu test system was developed \cite{wu_development_2019}. 

Simulation tools such as technology computer-assisted design (TCAD) simulation and Monte-Carlo charge transport were employed to increase the understanding of the detector and provide feedback for the designers realizing the prototypes. The Allpix$^2$ framework \cite{spannagel_allpix2:_2018} was developed to integrate the benefits of Geant4 \cite{allison_recent_2016}, charge transport algorithm and TCAD simulations into a user friendly platform for simulation of silicon detectors.

\subsection{The CLICdp Timepix3 telescope}

The Timepix3 ASIC \cite{poikela_timepix3:_2014} is a readout designed for silicon and gaseous pixelated detectors with an array of $256 \times 256 $ pixels with a pitch of $\pitch{55}{55}$, providing  for each detected hit a \SI{10}{bit} Time-Over-Threshold (TOT) measurement with \SI{25}{\nano\second} granularity and a \SI{14}{bit} Time-Of-Arrival (ToA) measurement with a \SI{1.5625}{\nano\second} binning. 

\begin{figure}
\begin{minipage}[b]{0.4\textwidth}
\begin{center}
    \includegraphics[width=\textwidth]{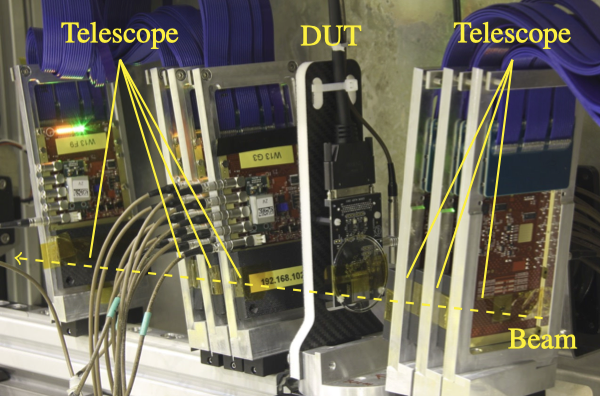}
    \caption{The CLIC Timepix3 telescope installation at CERN SPS H8 beamline.}
    \label{timepix3telescope}
    \end{center}
\end{minipage}
\hspace{0.1\textwidth}
\begin{minipage}[b]{0.4\textwidth}
    \begin{center}
    \includegraphics[width=0.8\textwidth]{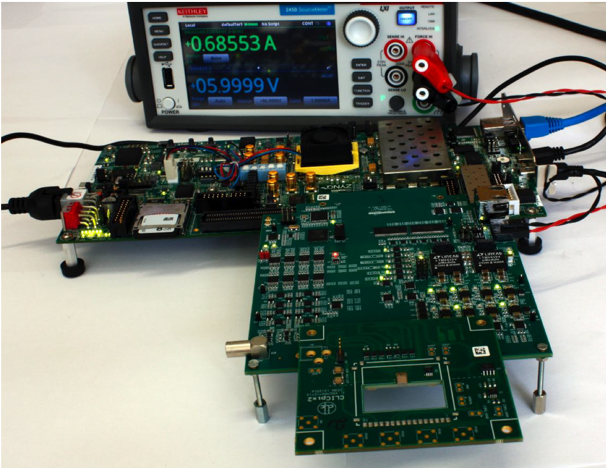}
    \caption{The CaRIBOu readout system hardware with a mounted CLICpix2 carrier board. }
    \label{caribou}      
    \end{center}
\end{minipage}
\end{figure}

The telescope consists of $7 \times$ Timepix3 assemblies readout by the SPIDR system \cite{visser_spidr:_2015}. The clock distribution to the planes is assured by a Trigger Logic Unit (TLU) providing the time reference ($t_0$) and measuring accurately the coincidence signal of two scintillator tiles located at each end of the telescope. The mechanical assembly of the telescope is illustrated in Figure \ref{timepix3telescope}. 

The Devices Under Test (DUTs) characterized with the telescope are positioned at the center of the telescope and aligned using motorized rotation and translation stages. The clock from the TLU and the $t_0$ signal are provided to the DUT along with a trigger signal produced from the coincidence of the scintillator tiles. The data collected are reconstructed using the Corryvreckan and EUTelescope reconstruction framework \cite{daniel_hynds_corryvreckan_nodate,rubinskiy_eutelescope._nodate}. Combining the timing information of the scintillators and the telescope planes, a timing resolution per track of \SI{1}{\nano\second} is achieved along with a pointing resolution at the DUT of \SI{3}{\micro\meter}.

\subsection{The CaRIBOu test system}

\begin{wrapfigure}{R}{0.4\textwidth}
    \centering
      \vspace{-10pt} 
    \includegraphics[width=0.38\textwidth]{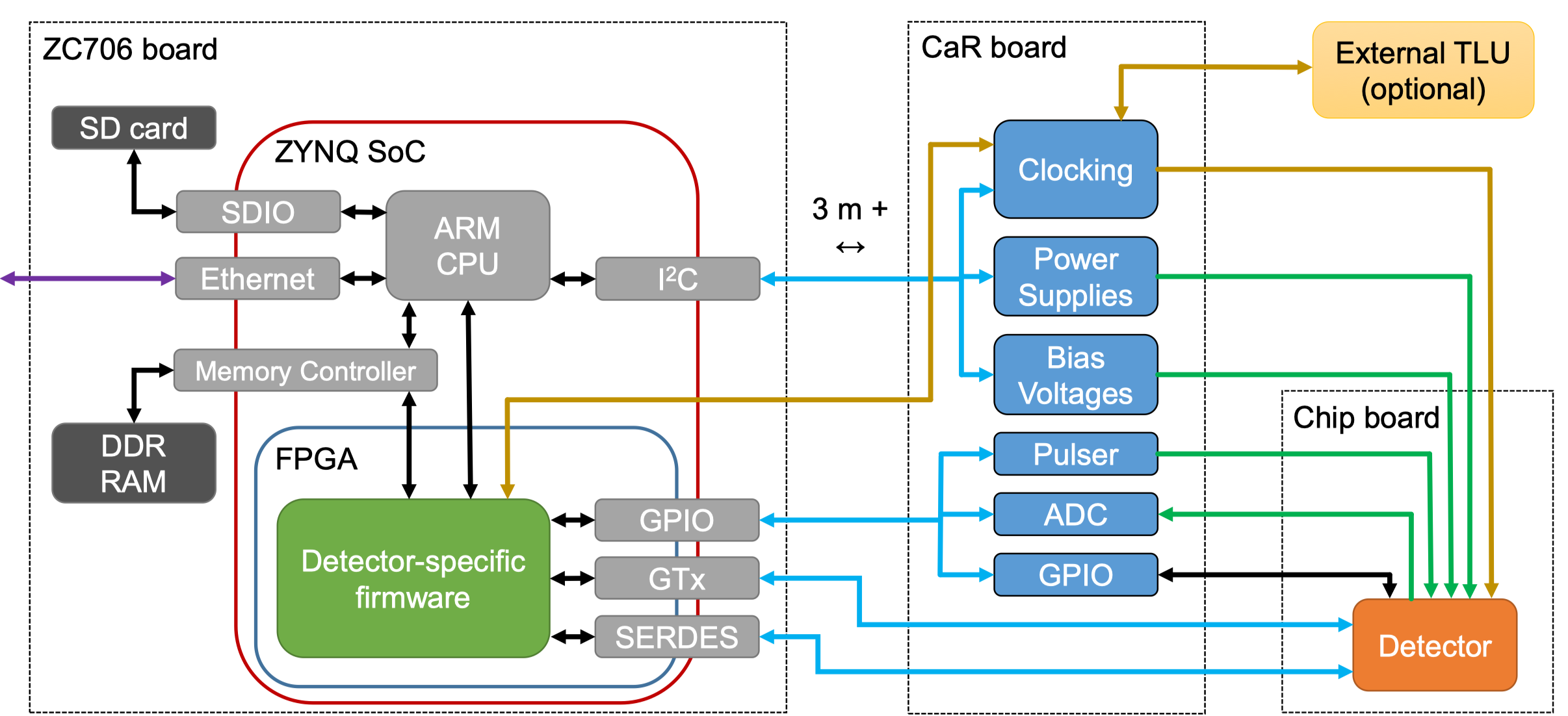}
    \caption{Schematics of the hardware, firmware and software interface of the CaRIBOu system.}
    \label{caribouscheme}
      \vspace{-20pt} 
\end{wrapfigure}

The control and readout interface board unit (CaRIBOu) \cite{wu_development_2019} is a versatile readout system designed to facilitate the design of the interface to pixel detector prototypes. It consists of a hardware system, composed of the CaR board, the Zynq ZC706 FPGA development board and a prototype specific carrier board unit, as illustrated in Figure \ref{caribou}. 

The CaR board is a custom board interfacing to the Zynq FPGA that provides the resources needed for a variety of prototypes, listed below.

\begin{itemize}
  \setlength\itemsep{-0.5em}
    \item $8~\times$ adjustable power supplies, \SIrange{0.8}{3.6}{\V},\SI{3}{\A}
    \item $32~\times$ adjustable voltage references, \SIrange{0}{4}{\V}
    \item $8~\times$ adjustable current references, \SIrange{0}{1}{\milli\A}
    \item $4~\times$ programmable injection pulsers
    \item $10~\times$ output and $14~\times$ input CMOS signal, \SIrange{0.8}{3.6}{\V}
    \item $17 \times$ LVDS signal routed to FPGA + $8\times$ full-duplex GTx links for high speed data transmission up to \SI{12}{\giga bit\per \s}
    \item Low jitter clock generator, slow and fast ADC, $8~\times$ \SI{50}{\kilo s \per \s}, $16~\times$ \SI{65}{\mega s \per\s}
\end{itemize}

The firmware for the CaRIBOu system is developed using a modular architecture with the different functionalities needed for the readout integrated into self-contained Intellectual Property blocks (IPs) interfaced to the Zynq ARM CPU through a standard AXI4 bus. The software for the system is based on the yocto Linux distribution and provides the libraries to control the interfaces connected to the FPGA, as illustrated in Figure \ref{caribouscheme}. The software provides a Hardware Abstraction Layer (HAL), a convenient way to control all the resources of the CaR board and communicate with the device specific firmware IP blocks implemented for each DUT. For test-beam operation with the Timepix3 telescope, an interface to the TLU receiving the clock and $t_0$ signal have been implemented to synchronize the DUTs to the telescope and enable their characterization.

\subsection{TCAD and Allpix$^2$ Monte-Carlo simulation}

During the detector design phase of our prototypes, input from TCAD simulation is used to provide guidance to make technological choices. However, the computing time required for simulating the interaction of particles with the detector spans from minutes for simplified two dimensional simulations to hours for more complex three dimensional simulations which makes it impractical to predict the behavior of devices in statistical terms. 

\begin{figure}
\centering
    \begin{minipage}[t]{0.35\textwidth}
        \begin{center}
        \includegraphics[width=0.85\textwidth]{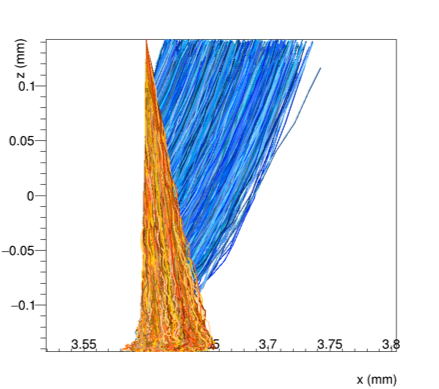}
        \caption{Propagation of the generated electrons (blue) and holes (red) in the substrate of an example pixel detector.}
        \label{propagation}
        \end{center}
    \end{minipage}
    \hspace{3cm}
    \begin{minipage}[t]{0.3\textwidth}
        \begin{center}
        \includegraphics[width=0.8\textwidth]{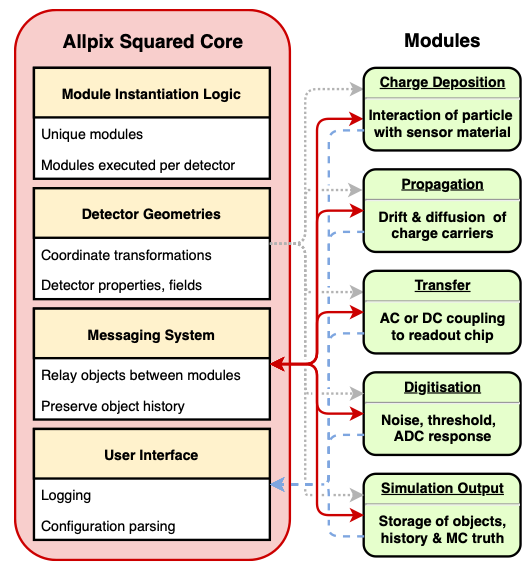}
        \caption{Structure of the Allpix$^2$ framework for a typical simulation.}
        \label{allpixscheme}
    \end{center}
    \end{minipage}
\end{figure}

The Allpix$^2$ framework \cite{spannagel_allpix2:_2018} was developed to provide a versatile, generic tool for Monte-Carlo simulations of pixel detectors. The geometry to be simulated is described using text interface and includes all the features of pixel detectors such as support  printed circuit boards (PCB), bump-bonds and other materials attached to the device. The particle sources are defined and the interactions of the particles with the detectors are simulated using Geant4. The resulting deposited energy, converted to electron-hole pairs, can be propagated through the bulk of the sensors using the provided propagation modules that include effects of drift, diffusion and the Lorentz effect in presence of a magnetic field as shown in Figure \ref{propagation}. The electric field can be selected from implemented models or imported from a TCAD simulation software in the DF-ISE format. The resulting propagated charges can then be converted to signal provided to the front-end electronics. The digitization of the signal is performed using parametrized models of the front-end response. The final simulation results can be output in various telescope reconstruction frameworks such as Corryvreckan \cite{daniel_hynds_corryvreckan_nodate}, Proteus \cite{mcgoldrick_synchronized_2014} and Eutelescope \cite{rubinskiy_eutelescope._nodate}. This feature allows for easy comparison between simulated and test-beam data. 

The software structure of Allpix$^2$ is highly modular, with a core software providing the required facilities such as geometry description, parsing tools, logging tools and persistent storage of the generated data. The simulation steps, from energy deposition to digitization are enclosed in modules that make use of the core functions. New modules can easily be integrated to replace or supplement the simulation flow. A typical simulation configuration is illustrated in Figure \ref{allpixscheme}.

\section{Vertex and tracker detector prototypes}

The CLICdp collaboration's effort in developing new technologies for vertexing and tracking covers the study of hybrid solutions and novel monolithic pixel detector technologies. Hybrid detectors allow for more complex integration and logic by separating the readout electronics from the sensor, at the cost of more complicated interconnect technology. Planar sensors, capacitively-coupled CMOS sensors (CCPD) and enhanced lateral drift sensors (ELAD) are under study. The monolithic CMOS approach, integrates the sensor and readout in the same silicon die, therefore it avoids the complex interconnect process and achieve lower material budget, at the cost of reduced functionality with regard to the hybrid solution. Silicon-on-insulator (SOI) CMOS pixels, high-voltage HV-CMOS and high-resistivity HR-CMOS technologies are under study. SOI-CMOS and Enhanced lateral Drift (ELAD) hybrid sensors are covered in a separate proceeding article of this series. 

\subsection{Planar sensor assemblies and CLICpix2}

\begin{figure}
\begin{minipage}[b]{0.5\textwidth}
    \captionof{table}{CLICpix2 in-pixel analog front-end performance for continuous operation.}
    \includegraphics[width=0.8\textwidth]{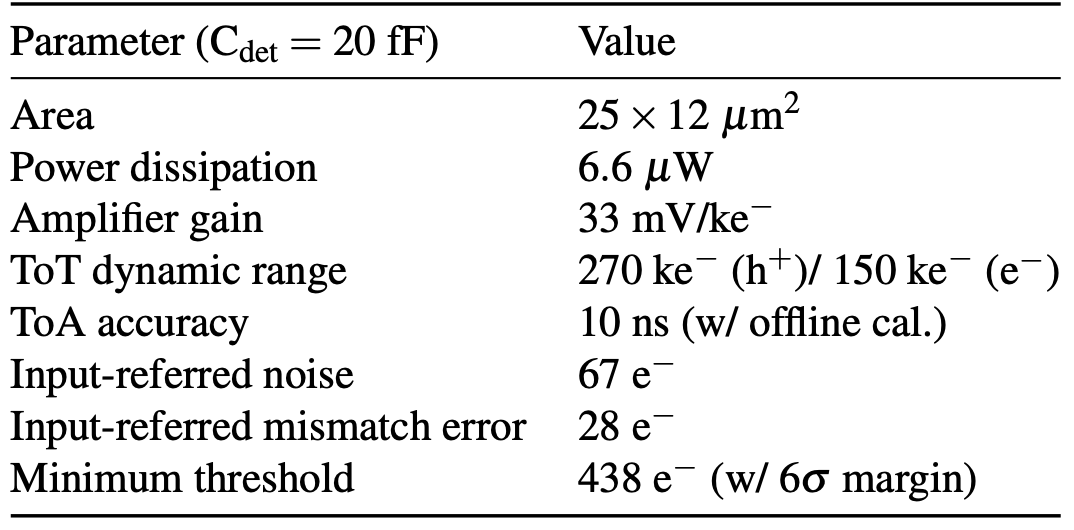}
    \label{clicpix2specs}
\end{minipage}
\hspace{1cm}
\begin{minipage}[b]{0.4\textwidth}
\centering
    \includegraphics[width=0.8\textwidth]{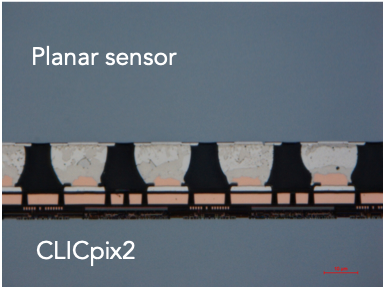}
    \caption{Bump bonds in a cross-section of a CLICpix2 and planar sensor assembly. Credits: IZM}
    \label{bump}
    \end{minipage}
\end{figure}

The CLICpix2 ASIC \cite{e._santin_clicpix2_2016} is a \SI{65}{\nano\meter} CMOS pixel readout designed to meet the requirements of the CLIC vertex detector. The matrix consists of 128 $\times$ 128 pixels with a pitch of $\pitch{25}{25}$. Each pixels provide a \SI{5}{bit} TOA and \SI{8}{bit} TOT measurement over a clock up to \SI{100}{\mega\hertz}. Power-pulsing of the matrix and readout circuitry is implemented and executed through an external control signal to reach the power consumption target for CLIC vertex detector of \SI{50}{\milli\watt\per\cm\squared} in CLIC operation conditions. Table \ref{clicpix2specs} summarizes the achievable simulated performance of the analog front-end. A low threshold ($\geq \SI{440}{e}$) and low noise (\SI{70}{e}) are required to handle the small signal of the thin, low material budget sensors under study. 

Hybridization of the CLICpix2 to planar sensors produced with active-edge technology produced with Advacam and FBK was performed at IZM using \chem{SnAg} bump-bonding technology. The ASICs, already diced, were mounted on support wafers and went through the bump deposition process. The fine pitch of \SI{25}{\micro\meter} of the ASIC and sensor represents a challenge for this technology, but preliminary results show that interconnect yields of above 99.9\% can be achieved. Figure \ref{bump} shows an example of successful bump connections between the sensor and ASIC. Further studies of the assemblies are ongoing and new high-density interconnect technologies such as an Anisotropic Conductive Films (ACF) are explored.

\subsection{Capacitively Coupled Pixel Detector (CCPD) sensor assemblies}

\begin{figure}
    \centering
    \includegraphics[width=0.315\textwidth]{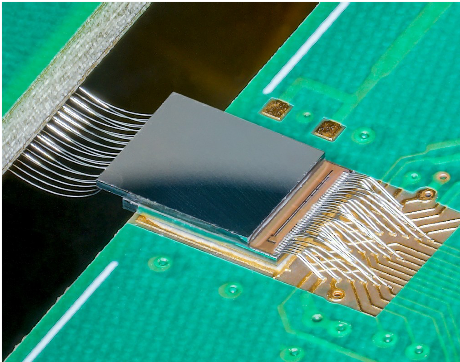}
    \includegraphics[width=0.6\textwidth]{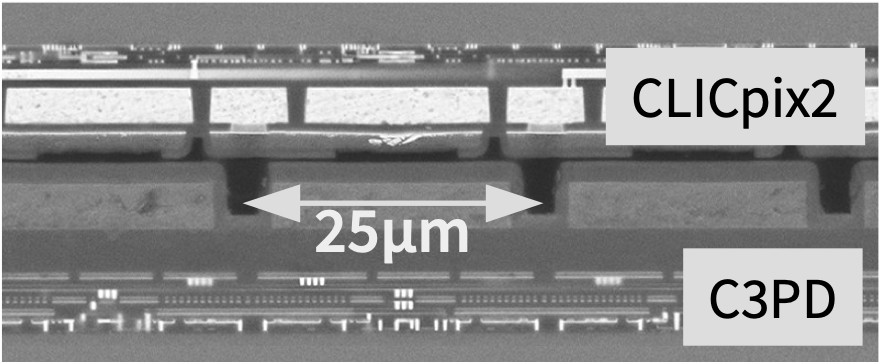}
    \caption{C3PD and CLICpix2 assembly wire-bonded to its carrier board (left), electron microscopy of the C3PD and CLICpix2 ASIC interface after gluing. A layer of less than \SI{100}{\nano\meter} of glue is present between the passivation layers, creating a large capacitance for the transmission of the signal (right).}
    \label{CCPDglueing}
\end{figure}

CCPDs were investigated using sensors designed for AC coupling to readout ASIC and produced in a commercial \SI{180}{\nano\meter} HV-CMOS technology. The C3PD sensor \cite{kremastiotis_design_2017} and its predecessor the CCPDv3 \cite{tehrani_capacitively_2016} were produced with the footprint of the CLICpix2 and its predecessor the CLICpix \cite{valerio_prototype_2014}. The coupling to the ASIC is performed through a thin glue layer applied to the sensor with a glue dispenser, followed by precision alignment and connection with the use of an ACC$\mu$RA 100 high accuracy bonder \cite{noauthor_accura_nodate}. Figure \ref{CCPDglueing} shows an example of a glue assembly on a PCB and of the achieved interface between ASIC and sensor. 

\begin{figure}
\centering
\subfloat[Most Probable Value (MPV) for energy deposition in TOT unit versus bias voltage.]{\includegraphics[width=0.31\textwidth]{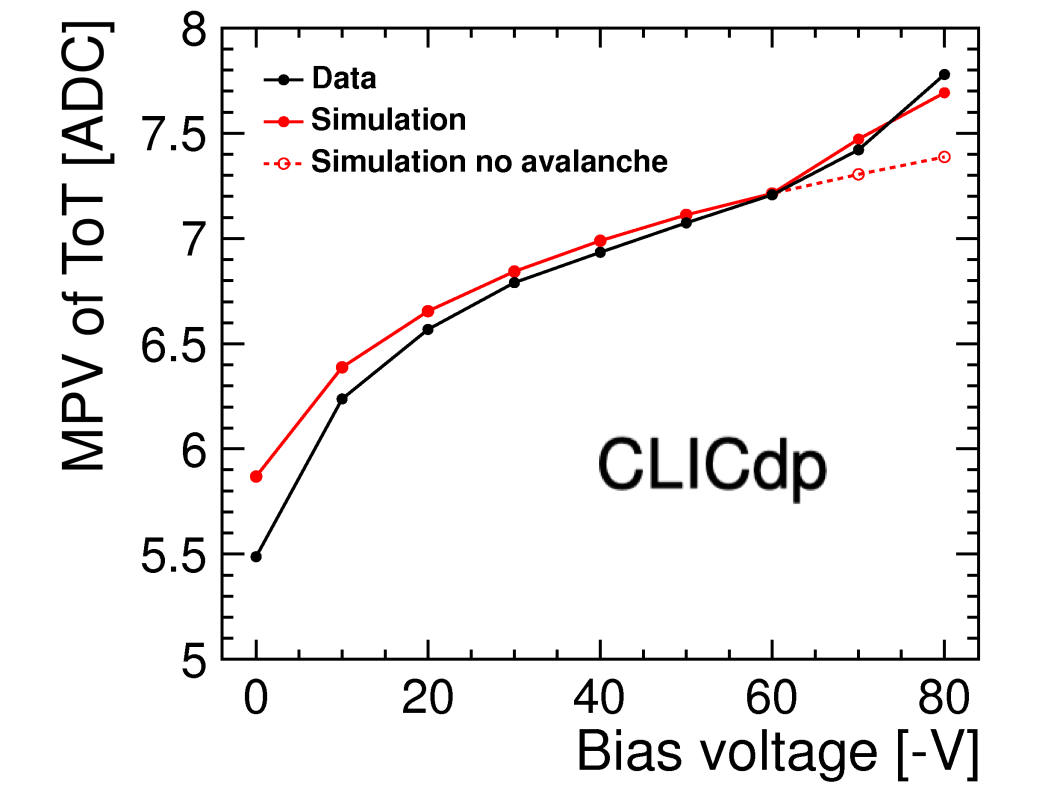}}
\hspace{0.01\textwidth}
\subfloat[MPV versus incident particle angle.]{\includegraphics[width=0.31\textwidth]{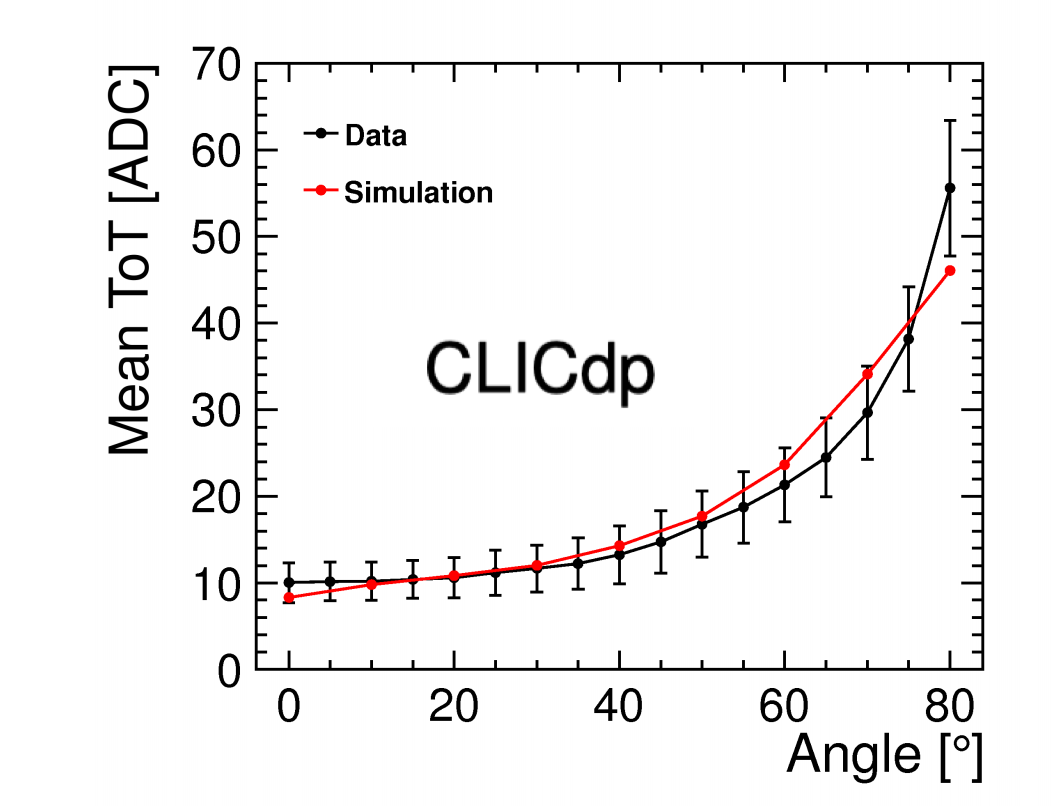}}
\hspace{0.01\textwidth}
\subfloat[Cluster size versus incident particle angle.]{\includegraphics[width=0.31\textwidth]{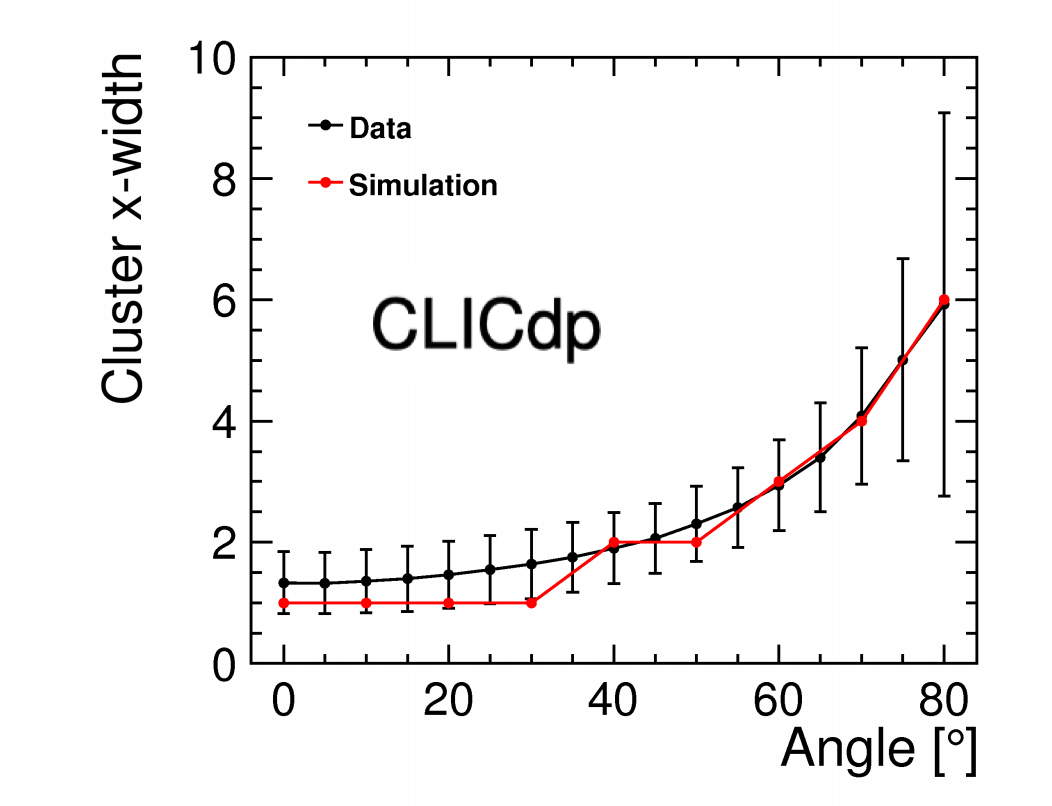}}%
\caption{Comparison of test-beam data and TCAD simulation for the CCPDv3 sensor coupled to the CLICpix ASIC \cite{buckland_simulation_2018}.}
\label{ccpd_sim_tb}
\end{figure}

The CCPDv3 and C3PD were simulated using TCAD \cite{buckland_simulation_2018}. The signal generated by particles at different incident angle was simulated and the transfer function of the readout electronics was applied to the simulated signal. The produced assemblies were characterized in a test-beam using the Timepix3 telescope. Figure \ref{ccpd_sim_tb} shows a comparison of results obtained with TCAD and front-end simulation and the experimental results obtained from test-beam. Further characterization of CCPD assemblies can be found elsewhere \cite{alipourtehrani_tracking_2019}. 

\subsection{Monolithic small fill-factor sensors}

\begin{figure}
\centering
    \subfloat[Reconstructed charge.]{\includegraphics[width=0.28\textwidth]{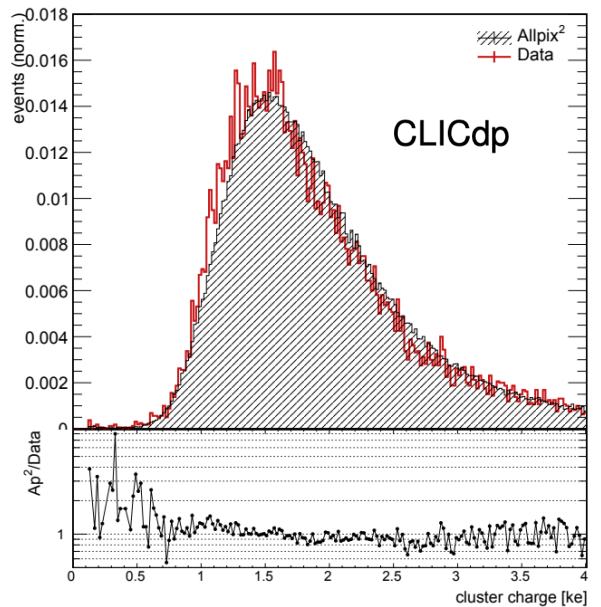}}
\hspace{0.01\textwidth}
    \subfloat[Mean cluster size vs. threshold.]{\includegraphics[width=0.28\textwidth]{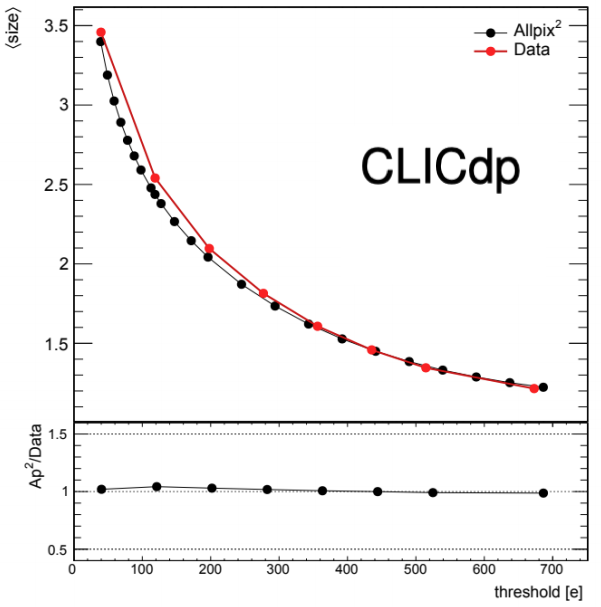}}
\hspace{0.01\textwidth}
    \subfloat[Single-point resolution vs. threshold.]{\includegraphics[width=0.28\textwidth]{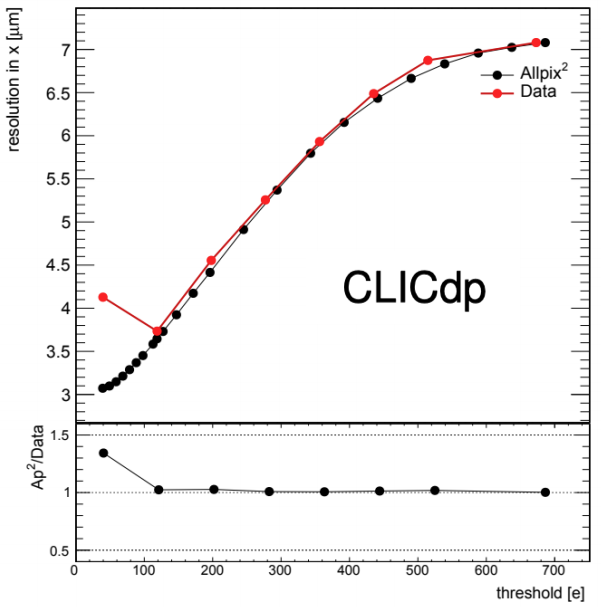}}
    \caption{Comparison of measured and simulated tracking characteristics of the ALICE Investigator}
    \label{investigator_tb}
\end{figure}

For the CLIC tracking detector,  monolithic pixel sensors are foreseen due to their large scale production capabilities and low material budget achievable. Small fill-factor CMOS sensors are composed of a small  n-type collection electrode separated from a large p-type well in which the readout electronics is located. The separation of the CMOS circuitry from the collection diode well allows reducing cross-talk between the two elements. The small size of the collection electrode leads to a small input capacitance and therefore low noise and power consumption of the front-end.  The ALICE Investigator chip \cite{munker_study_2017} was characterized in a test-beam using the Timepix3 telescope. Detailed TCAD simulations of the different diode geometries included in the Investigator were produced \cite{munker_test_2018} in order to understand the charge collection process with small collection diodes. The electric field simulated was imported into Allpix$^2$ and a detailed simulation of the test-beam was carried out. Figure \ref{investigator_tb} show comparisons of the simulation to test-beam measurements showing a good agreement between the model and collected data.

\begin{wrapfigure}{r}{0.4\textwidth}
\includegraphics[width=0.38\textwidth]{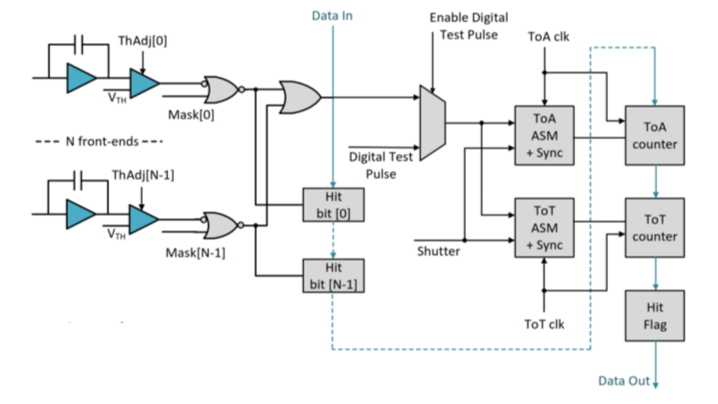}
\caption{Schematics of the CLICTD pixel front-end.}
\label{clictd}
\end{wrapfigure}

Following the encouraging results from the ALICE Investigator, a CLIC specific sensor has been designed meeting the requirements of the CLIC tracking detector. Using TCAD and Monte-Carlo simulation validated with Investigator data, the implant layout and geometry of the pixels was optimized to meet CLIC requirements. Details of the simulations can be found elsewhere in the proceeding series. The CLICTD technology demonstrator consists of physical pixels of $\pitch{30}{37.5}$ containing a pre-amplifier and discriminator combined in a logical pixel of $\pitch{30}{300}$, as illustrated in Figure \ref{clictd}. Each logical pixel can measure TOT with \SI{5}{bit} precision and TOA with \SI{8}{bit} precision using a \SI{100}{\mega\hertz} clock. Each logical pixel also provides the information on which physical pixels were fired during the active period. The ASIC implements power-pulsing of the front-end and readout electronics in order to meet CLIC power consumption requirements. CLICTD was sent for production in February 2019. 

\subsection{Monolithic large fill-factor sensors}

Large fill-factor Monolithic CMOS sensors implement the readout electronics in a large n-type well that acts as the collecting electrode. The ATLASpix simple sensor \cite{peric_high-voltage_2018}, originally designed for the ATLAS CMOS collaboration, is being characterized in a test-beam to evaluate its performance with regard to CLIC tracking detector specifications.  It consists of a large matrix of \SI{16} \times \SI{3.25}{\milli\meter\squared} pixels with a pitch of $\pitch{130}{40}$. Each pixel provides a \SI{10}{bit} TOA and \SI{6}{bit} TOT measurement on a clock of up to \SI{160}{\mega\hertz}. Data are readout through a \SI{1.6}{\giga bit \per\s} serial link. A prototype of the ATLASpix sensor with a substrate resistivity of \SI{200}{\ohm\cdot\cm} was thinned to a thickness of \SI{100}{\micro\meter} and characterized in a test-beam using the Timepix3 telescope. 

\begin{figure}
\centering
    \subfloat[Detection efficiency versus bias voltage at the nominal threshold of 1000e.]{\includegraphics[width=0.59\textwidth]{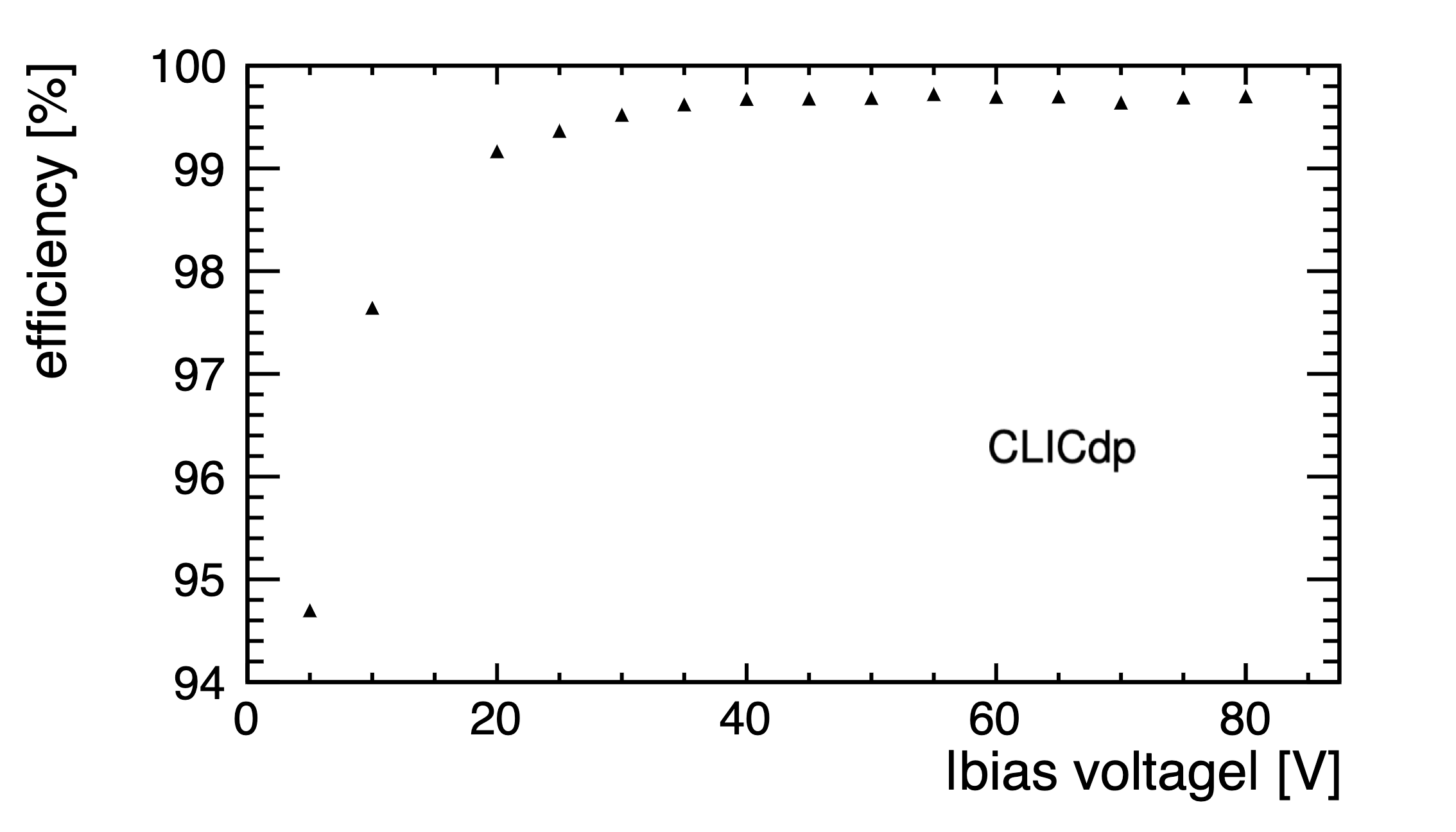}}
    \subfloat[Timing distribution after correction.]{\includegraphics[width=0.41\textwidth]{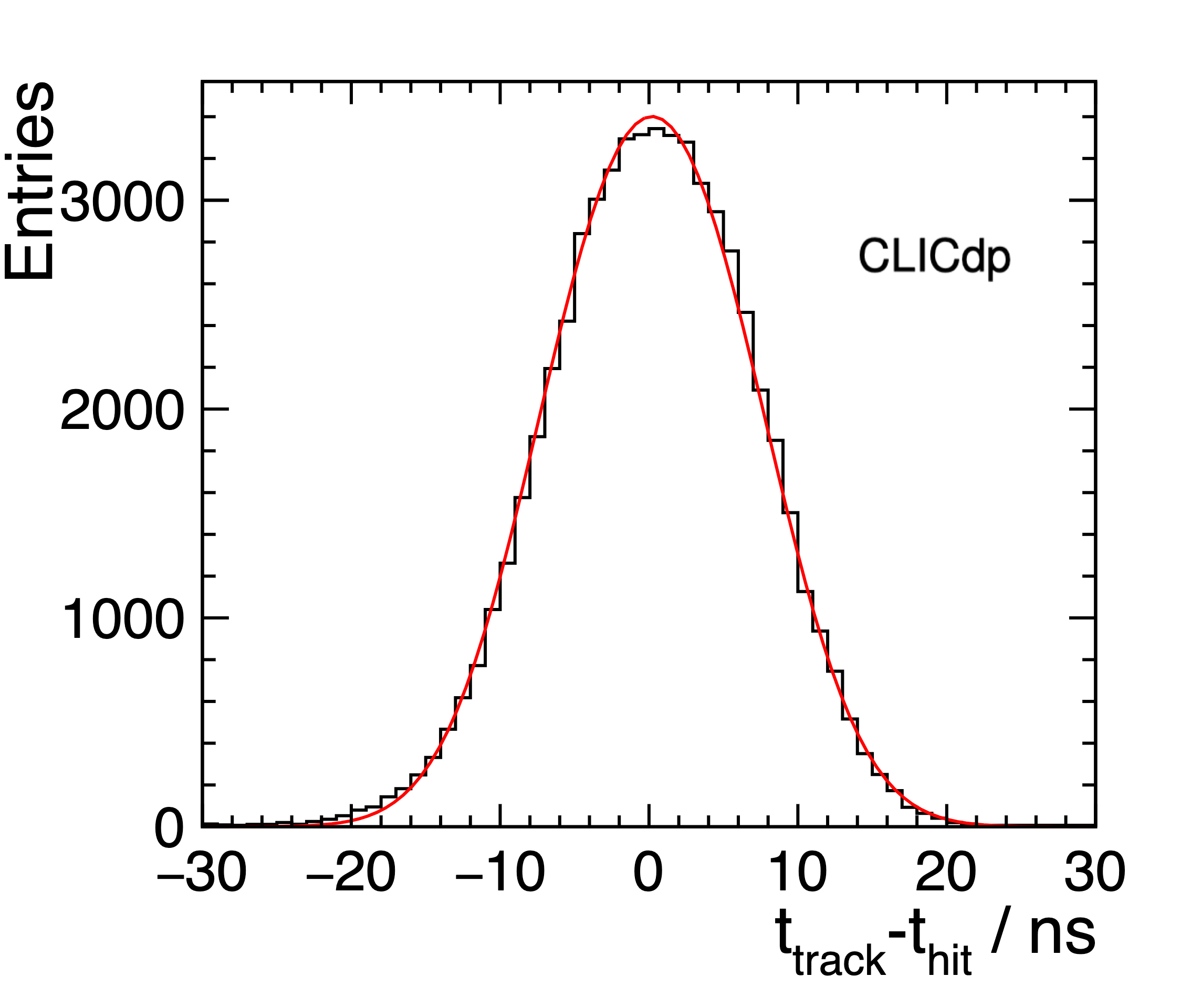}}
    \caption{ATLASpix test-beam characterization results.}
    \label{ATLASpix_tb}
\end{figure}

Figure \ref{ATLASpix_tb} shows the results of the detection efficiency, spatial and temporal resolution of the ATLASpix prototype. Due to the low resistivity and thickness of the substrate, only a low amount of charge sharing was observed (approximately 10\% multi-pixel clusters) and the spatial resolution in both directions is compatible with the pitch of the pixel. Good detection efficiency at perpendicular incidence angle was observed over a large range of bias voltages. The timing resolution was measured using a \SI{16}{\nano\second} binning. Correction for a time-walk and for the systematic delay of each row of the sensor was applied offline. A timing resolution $\sigma_t$ of \SI{7.2}{\nano\second} (RMS) is achieved. Taking into account the TOA binning and the lack of time structure in the CERN SPS beam, a fit of a box function convoluted to a Gaussian function was performed. For a fixed box width corresponding to the TOA binning, an intrinsic resolution for the pre-amplifier and discriminator of \SI{5.6}{\nano\second} can be extracted. 

\section{Conclusion}

The CLICdp vertex and tracker R\&D focuses on identifying the most promising technologies for the realization of a detector meeting CLIC physics requirements. The CLICdp Timepix3 telescope and the CARiBOu readout system were developed to efficiently perform characterization of the prototypes in test-beams and in laboratory. TCAD simulations, in combination with the Allpix$^2$ framework, have been developed and used to gain understanding of current and future prototypes. The CLICpix2 ASIC was developed, meeting CLIC specifications, and studied using planar and CCPD sensors developed by the collaboration. Monolithic sensors with small and large fill-factor electrodes were simulated and characterized and the obtained results show good agreement with the simulation. Further prototypes fulfilling  further the CLIC vertex and tracker requirements are designed and being produced.

\acknowledgments
This project has received funding from the European Union’s Horizon 2020 research and innovation programme under grant agreement No 654168.

\printbibliography

\end{document}